\documentclass[conference]{IEEEtran}
\IEEEoverridecommandlockouts
\usepackage{cite}
\usepackage{graphicx}
\usepackage{amsmath}
\usepackage{subfigure}
\usepackage{gensymb}
\usepackage{xcolor}

\begin{document}
\title{Effect of Standing Wave on Terahertz Channel Model\\
\thanks{This work was funded by the NSF under awards CNS-1618936 and CNS-1910655.}
}

\author{\IEEEauthorblockN{Ha Tran}
\IEEEauthorblockA{Department of ECE\\
Portland State University\\
Portland, OR 97207\\
Email: tranha@pdx.edu}
\and
\IEEEauthorblockN{Thanh Le}
\IEEEauthorblockA{Department of ECE\\
Portland State University\\
Portland, OR 97207\\
Email: thanh4@pdx.edu}
\and
\IEEEauthorblockN{Suresh Singh}
\IEEEauthorblockA{Department of Computer Science\\
Portland State University\\
Portland, OR 97207\\
Email: singh@cs.pdx.edu}}
%
%
%


\markboth{Terahertz Channel Model for High-gain Horn Antennas and Effect of Standing Wave}%
{}



\maketitle

\begin{abstract}
 There is a growing interest in exploiting the terahertz frequency band for future communication systems that demand high data rates. Given the complex propagation behavior of this frequency band, various researchers have developed channel models that can be utilized in the development of communication systems. These models however do not include a crucial aspect of terahertz propagation at short distances -- the presence of standing waves. Our measurements show that at specific distances, the effect of standing waves is significant. In this paper, we extend previous terahertz channel models to include the effect of standing waves and show a good fit with our measurements.
 Our measurements and modeling cover the five most promising terahertz frequency bands -- 140, 220, 340, 410, 460 GHz.
\end{abstract}


\section{Introduction}
It is expected that wireless data needs will increase
by three orders of magnitude over the next decade \cite{hemadeh18cst}, leading to an intense interest in exploiting the largely unused terahertz band for future wireless needs. As \cite{xing19globecom} notes, terahertz frequencies can be deployed to support new massively broadband IoT for sensing and communications as well as enable the deployment of wireless fiber links to rural communities. However, this frequency band has several challenges including poor propagation, environmental impairments and a lack of devices that can work across ultra-wide bandwidths. As a result, there are several efforts underway aimed at examining different aspects of this frequency band with the goal of providing high data-rate communication channels.

One set of research approaches has focused on demonstrating high speed communication at specific sub-bands while another approach has combined channel measurements with modeling to predict the achievable data rates. For example, data rates of tens of gigabit per second  (Gbps) for line of sight (LoS) and of several Gbps for non line of sight (NLoS) has been demonstrated at the 300-350 GHz frequency band \cite{kleine12cpem}. Hirata et al \cite{Hirata2010}
demonstrated 10 Gbps transmission at 120 GHz frequency over a distance
of almost 6 km. Kallfass et al \cite{Kallfass2011} demonstrated
transmissions using OOK at 220 GHz and achieved 50 Gbps over 50 cm
distances. Song et al \cite{Song2009} showed transmissions of 8 Gbps
using ASK modulation over 250 GHz. Most recently, Jia et al
\cite{jia18jlt} demonstrated 106 Gbps at 400 GHz using a photonic
wireless link. There have been many other demonstrations as well
\cite{Antes2011,Zhang2012,Song2012,Ducournau2014,Wang2013,Lu2013}
though a notable one was by Koenig et al \cite{Koenig2013} in which
they achieved 100 Gbps over 20 m using a phased antenna array. 

Other works on examining the achievable data rates rely on channel measurement and modeling approaches. Using this method, \cite{ma18apl} shows data rates of 1 Gbps are obtainable using 1 GHz wide frequency bands at 100, 200, 300, and 400 GHz. \cite{moshir16ncn} shows that data rates of several Tbps (terabit/sec) are achievable utilizing acousto-optic modulation (AOM) over the entire frequency band in conjunction with careful rate adaptation algorithms.

A consequence of this flurry of work on terahertz over the past decade has been a better understanding of how these signals propagate in indoor and outdoor environments. One of the key observations \cite{xing19globecom} is that terahertz propagation differs significantly from millimeter wave in important ways including, higher phase noise and Doppler, significant diffraction effects as frequencies increase \cite{kim15eucap}, and the presence of multiple arriving paths even without scatterers between the transmitter and receiver due to reflections from objects behind the transmitter and receiver. Furthermore, given the need for high-directivity antennas, we see somewhat different scattering patterns at terahertz as compared to millimeter wave \cite{kim16twc}.

Over the past year, we have been conducting measurements of a LoS channel using high directivity horn antennas \cite{singh-thz19globecom,singh20wcnc} at 140, 220, 340, 410, 460 GHz. These frequencies are identified as good Terahertz communication windows above 100GHz \cite{john10jap}. In addition to the effects observed by others, we noted a significant impact of a {\em standing wave} that gets established between the transmitter and receiver antennas. The effect of this is that the total attenuation at some distances is lower than we would expect if using a strict distance based attenuation as is typically done. {\em In this paper, we incorporate a model for the standing wave and develop an accurate channel model that correctly predicts the received signal for line of sight indoor channels.}

The remainder of the paper is organized as follows. In the next section, we present a brief overview of the current state of the art in terahertz channel models.  Section \ref{model} presents our channel model which is then compared against measurements in section \ref{validate}. In section \ref{testbed} we describe our measurement setup as well as the experimental design for the channel measurements. We conclude in section \ref{conclude}.

\section{Related Work} \label{related}

There has been considerable work on channel modeling for millimeter wave \cite{hemadeh18cst} and much of it is applicable to terahertz, with some modifications. Specific factors that need to be accounted for include frequency-dependent absorption due to water and other molecules, absence of many reflected paths due to material attenuation, consideration of high-gain antennas in propagation path estimation, frequency-dependent diffraction effects, presence of reflections between objects behind the transmitter and receiver (such as the body of the instrumentation), and, as we discovered, presence of {\em standing waves} at some distances.

One popular approach for building terahertz channel models is to utilize a Geometry-based Stochastic Channel Model (GSCM). In these models, ray-tracing is combined with geometrical properties of wave scattering objects modeled as random variables and an optics-based approach to model diffraction. Early ray-tracing path loss models for terahertz \cite{jansen08tap,priebe10imtw} were enhanced by combining frequency-dependent path gain \cite{molisch05tvt} with statistical models for multipath propagation \cite{saleh87jsac} giving us a much improved model \cite{priebe13twc}. The first GSCM model for terahertz was proposed in \cite{kim15eucap,kim16twc} for fixed point-to-point desktop communications. The model includes reflections that occur from behind the transmitter and receiver producing multiple paths to the receiver. \cite{han15twc} models frequency-based diffraction effects.

In our previous work \cite{moshir16ncn} we constructed and validated a LoS channel model that added a frequency-based absorption coefficient to the standard Friis equation. Measurements with a time-domain system showed good agreement. However, the system uses HDPE lenses for creating collimating beams and as a result effects of diffraction, reflections, and other impairments are absent. In subsequent studies, we used a frequency-domain system and horn antennas to conduct measurements \cite{singh-thz19globecom,singh20wcnc} for modeling MIMO channels. We observed that as distance increased, at some points the attenuation did not change. Investigation revealed the presence of standing waves that appear to have a significant impact on the channel. We investigate this behavior below.

\section{Channel Model} \label{model}

Our channel model is a modification of the {\em single-frequency floating intercept model} \cite{cheng17eucap}. In this model, the path loss (PL) is written as,
\begin{equation}
\mbox{PL}(d) = \alpha + 10\beta\log_{10}\left(\frac{d}{d_0}\right) + X_\sigma, \mbox{    } d \geq d_0
\label{sffl}
\end{equation}
where $d$ is the distance, $d_0 = 10$ cm is the reference distance, $\alpha$ is a floating intercept in dB that denotes the free-space path loss at $d_0$. The distance-based path loss exponent is $\beta$ and $X_\sigma$ is the large-scale shadow fading that is modeled as a zero-mean Gaussian with standard deviation $\sigma$. To estimate the parameters $\alpha, \beta, \sigma$ we use a least squares linear fit such that the root mean square deviation from the mean path loss is minimized.

In our measurements, we observed that the attenuation remains almost constant for some distances before increasing again. We realized that this behavior occurs because standing waves created between the transmit and receive horn antennas increase the received signal strength. We therefore add an additional term to the above equation to account for this. 

\subsection{Standing Wave Model}

We can write the received voltage due to the standing wave $V(d)$ at distance $d$ as consisting of two components -- the forward wave and the reflected wave \cite{pozar2011}:
\begin{equation}
\begin{array}{l}
V_f(d) = e^{-ik(d-d_0)} A\\
V_r(d) = \Gamma e^{ik(d-d_0)}A\\
\end{array}
\label{forward-reflected}
\end{equation}
for some complex amplitude $A$ (corresponding to the forward wave at $d_0$), $k$ is the wavenumber, and $\Gamma$ is the reflection coefficient. The superposition of these two waves gives,

\begin{equation}
\begin{array}{ll}
V_{net}(d) & = V_f(d) + V_r(d)\\
& = e^{-ik(d-d_0)}(1 + \Gamma e^{i2k(d-d_0)})A\\
\end{array}
\label{net}
\end{equation}
Therefore, the magnitude of the standing wave can be written as,
\begin{equation}
\begin{array}{lll}
|V_{net}(d)|^2 &= &V_{net}(d)V_{net}^*(d)\\
&= &e^{-ik(d-d_0)}(1+\Gamma e^{i2k(d-d_0)})A \\
&&e^{ik(d-d_0)}(1+\Gamma^* e^{-i2k(d-d_0)})A^*\\
&= &\left[1 + |\Gamma|^2 + 2\Re(\Gamma e^{i2k(d-d_0)})\right] |A|^2\\
\end{array}
\label{vnet-amplitude}
\end{equation}
Since this quantity is added to the signal at the receiver, we need to subtract it (in dB) from the path loss formula in equation \ref{sffl} to obtain the final model for path loss. 

\begin{figure}[th]
  \centerline
      {\includegraphics[width=3in]{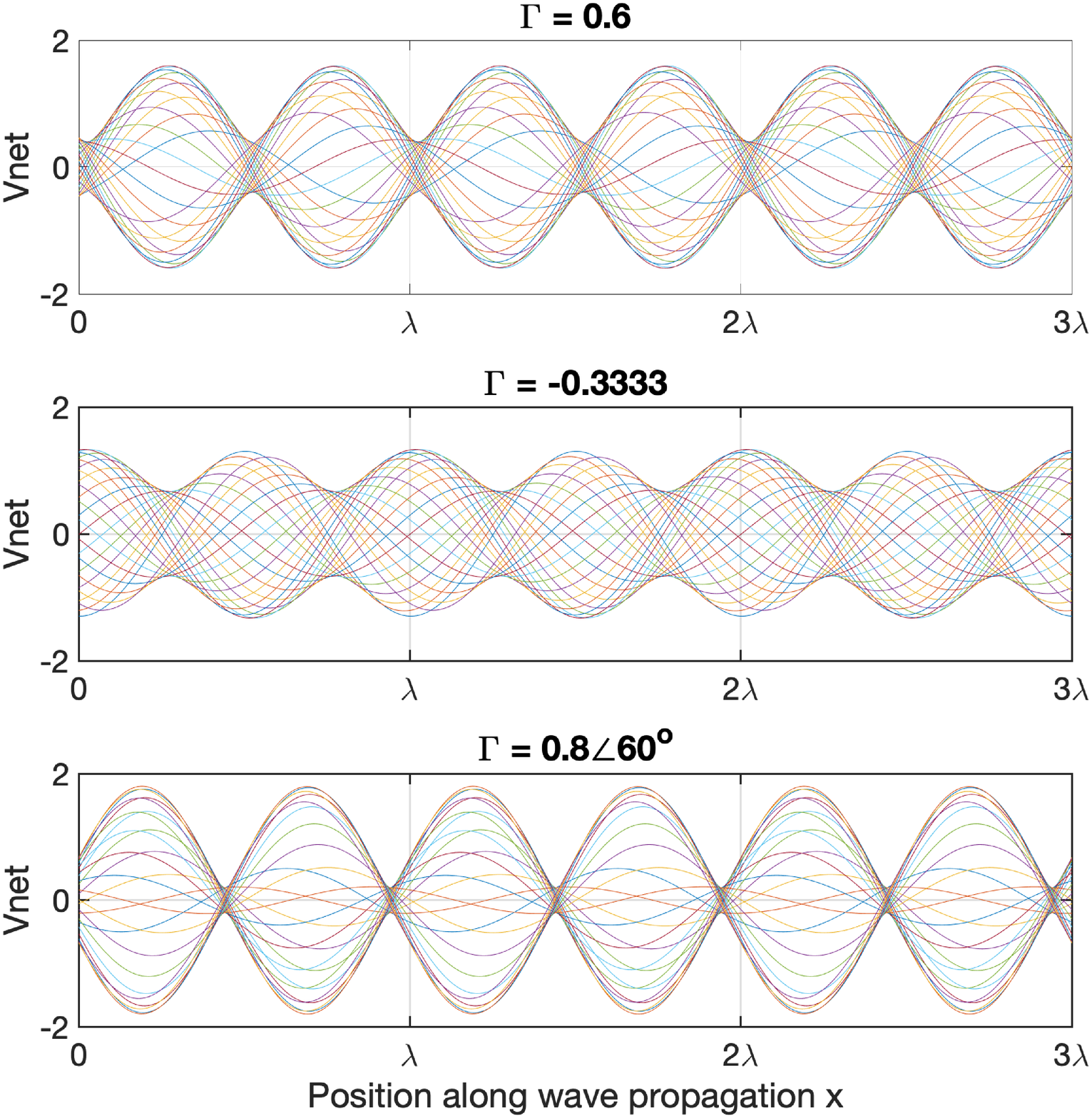}}
      \caption{Standing wave between the transmit and receive antennas.}
      \label{standingwave}
\end{figure}

\begin{figure}[th]
  \centerline
      {\includegraphics[width=2in]{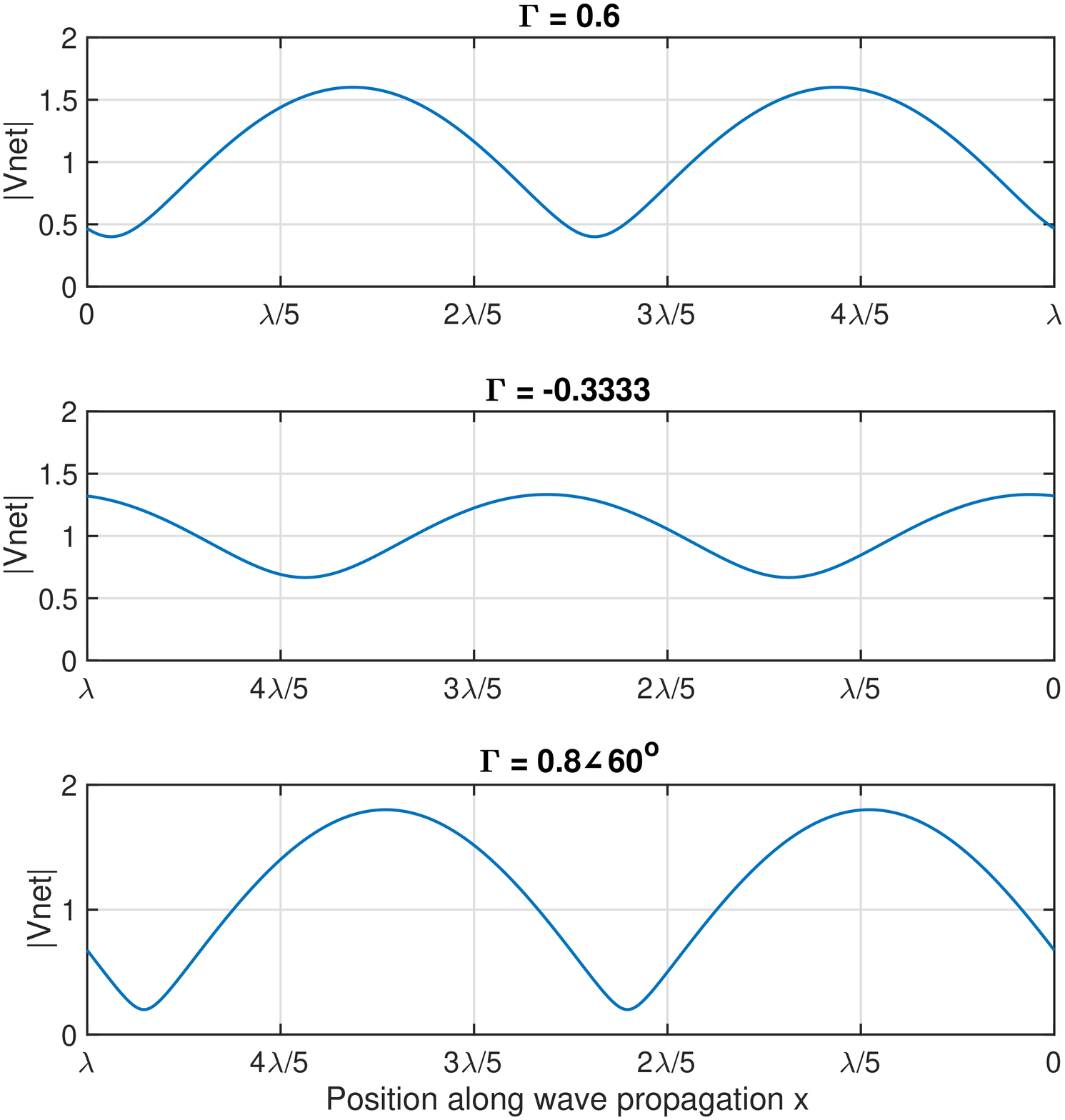}}
      \caption{Standing wave amplitude}
      \label{standingamplitude}
\end{figure}

In Figure \ref{standingwave} we plot the standing wave as a function of distance in term of $\lambda$ for three different values of the reflection coefficient $\Gamma$ = 0.6, -0.333, 0.8$\angle$60\degree. The standing wave ratio  (SWR) coefficient $V_{max}/V_{min}$  is 4, 2, 9 respectively. In Figure \ref{standingamplitude} we plot the amplitude of the standing wave for the three $\Gamma$ values for different distances starting with $d$ that is a multiple of $\lambda/2$ and then subsequent values reducing the distance by $\lambda/10$. 
From Figures 1 and 2, we can clearly see the effect of the standing wave on the received signal. Depending on the values of the complex reflection coefficient (its magnitude and phase), the magnitude of the standing wave can vary which leads to changes in the received signal.

\section{Measurement Setup and Experimental Design} \label{testbed}

\begin{figure}[th]
\centerline
{\includegraphics[width=2in]{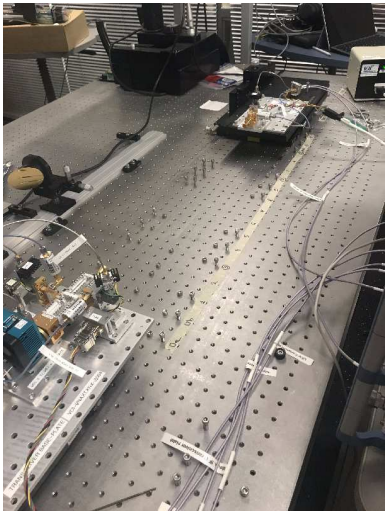}}
\caption{Measurement setup.}
\label{labsetup}
\end{figure}

We use a Rohde \& Schwartz Vector Network Analyzer (VNA) and Virginia Diodes, Inc. (VDI) frequency extender system to conduct the measurements. The system is 
capable of producing signals up to 700 GHz. Figure \ref{labsetup} illustrates the setup for frequency band WR-6.5 (110-170 GHz). For this band, the transmit and receive antennas we use are horn antennas with gain of 21 dB and a $13^0$ half power beam width (HBW).  Measurements were conducted in the lab where the ambient temperature
was 72F and relative humidity was 40\%. Other parameters of the
experiment are given in Table \ref{table1}.

\medskip
\begin{table}
\centering
\caption{Measurement parameters}
\begin{tabular}{|l|l|} \hline
Output power & 5 dbm\\
Center Frequency & 140, 220, 340, 410, 460 GHz\\
$\Delta f$ & 0.1 GHz\\
IF Bandwidth & 1 kHz\\
Averaging & 10\\ \hline
Distance & 10.16 - 81.28 cm in 5.08 cm\\
& increments\\ \hline
\end{tabular}
\label{table1}
\end{table}
\medskip

\begin{figure}[th]
\centering
\subfigure[3D antenna pattern for WR-6.5.]
{\includegraphics[width=3in]{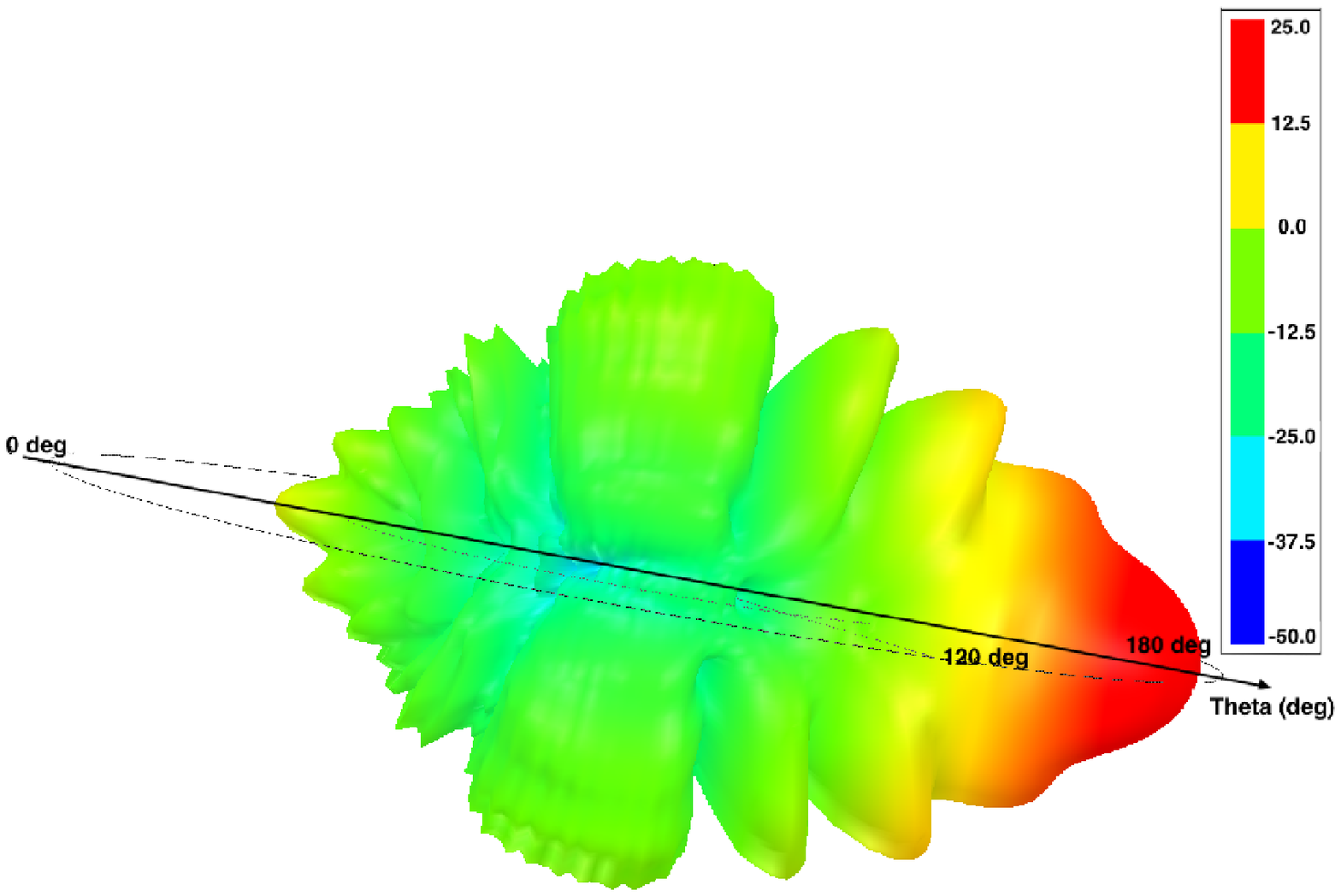}}
\subfigure[Angular antenna gain for different frequencies.]
{\includegraphics[width=3in]{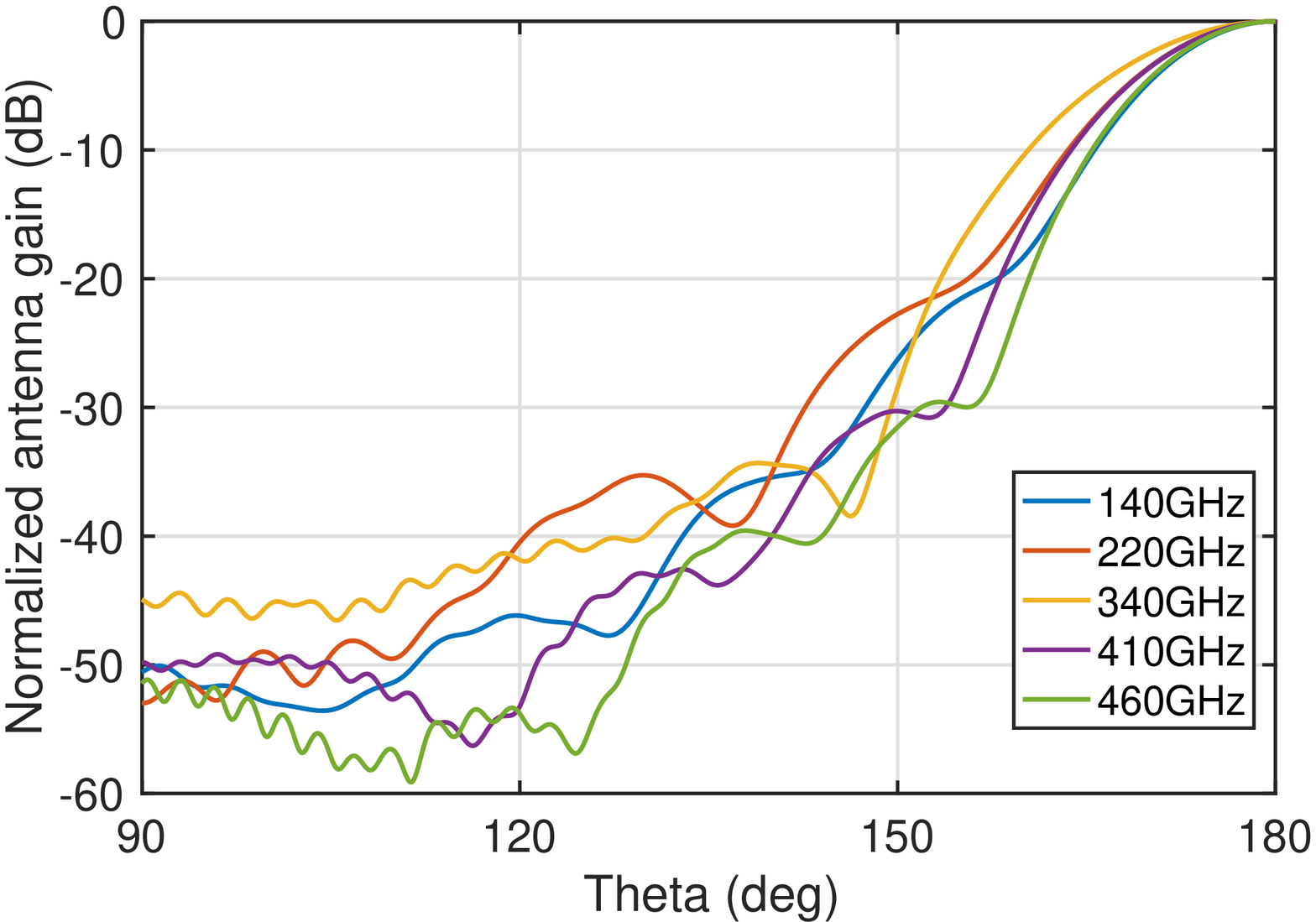}}
\caption{HFSS simulator output for antenna gain.}
\label{wr65}
\end{figure}

Notice that the horn antenna used for different frequency bands is different which also yields different received signals. The specifications of the VDI antennas for the measurements are given in Table \ref{table2}. Figure \ref{wr65}(a) demonstrates the 3D antenna pattern for WR-6.5 and the simulated antenna gain for investing frequencies are plotted in Figure \ref{wr65}(b).

\medskip
\begin{table}
\centering
\caption{VDI Horn antennas specifications}
\begin{tabular}{|l|l|l|l|l|} \hline
& Freq. Band & HBW & Gain & Beam Waist\\
&&&&Radius\\ \hline
WR-2.2 & 325-500 GHz & $12^0$ & 25 dB & 1.3 mm\\ \hline
WR-4.3 & 170-260 GHz & $13^0$ & 21 dB & 2.7 mm\\ \hline
WR-6.5 & 110-170 GHz & $13^0$ & 21 dB & 4.1 mm\\ \hline
\end{tabular}
\label{table2}
\end{table}

\section{Measurements and Validation} \label{validate} 

\begin{figure}[th]
\centerline
{\includegraphics[width=3in]{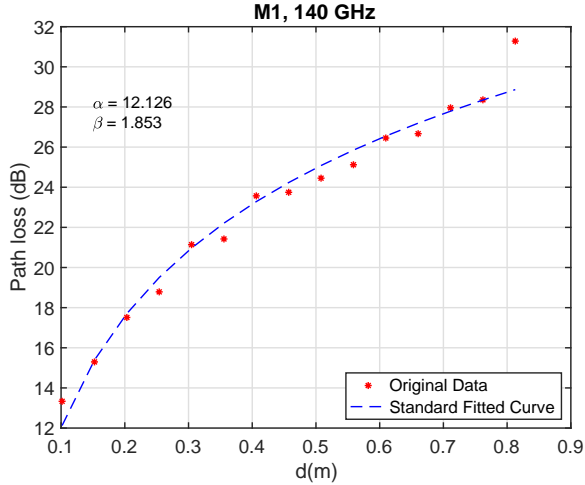}}
\caption{Equation \ref{sffl} rms fit to measured data for 140 GHz.}
\label{140-sffl}
\end{figure}

To explain the process of fitting our model to the measured data, let us initially focus on measurements conducted at 140 GHz for varying distances. In Figure \ref{140-sffl} we plot the measured data points and the {\em root mean square} (rms) fit for equation \ref{sffl} to the data. While the trend of the fit is in good agreement, we observe that the actual data shows a curious behavior. We can see that for pairs of distances, the attenuation is flat. This happens due to the effects of the standing wave which gets established between the transmitter and receiver. In the next subsection, we explain how the fit is improved by incorporating equation \ref{vnet-amplitude} into the overall model. Subsequently, in section \ref{linear-data}, we show the model fit for the other frequencies. 

\subsection{Parameter Fitting}

\begin{figure}[th]
\centerline
{\includegraphics[width=3in]{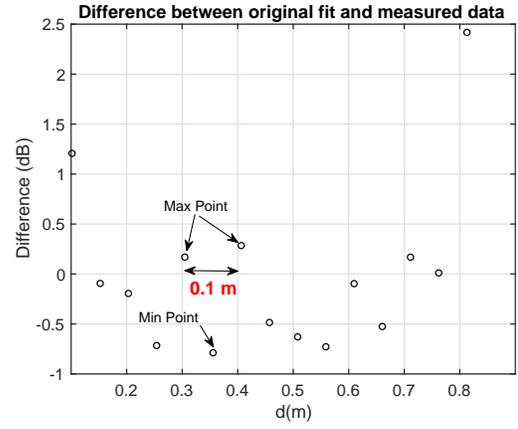}}
\caption{Difference of fit and measured for 140 GHz.}
\label{140-standing}
\end{figure}

Equation \ref{sffl} and \ref{vnet-amplitude} contain several parameters including $\alpha, \beta, \Gamma, k$ that we need to estimate from the measurements. Using the rms fit shown in Figure \ref{140-sffl} we obtain $\alpha, \beta$ (shown in the figure). To find the reflection coefficient $\Gamma$ and wavenumber $k$, we begin with the SWR,
\[
\mbox{SWR} = \frac{V_{max}}{V_{min}} = \frac{1+|\Gamma|}{1-|\Gamma|}
\] 
which gives us,
\[
|\Gamma| = \frac{V_{max}-V_{min}}{V_{max}+V_{min}}
\]
In order to measure $V_{max}$ and $V_{min}$, we utilize Figure \ref{140-standing} which is obtained by subtracting the fit in Figure \ref{140-sffl} from the shown measured data points. We observe maximum and minimum amplitudes (pointed out in the diagram). We utilize the largest and smallest values to the two maximum and minimum voltage levels. In our case, these values are 0.28 and -0.79 dB respectively. Using these values, we obtain $|\Gamma| = 0.158913332$ and the phase of $\Gamma$ is $0.25\pi$. The last parameter we need to estimate is the wavenumber $k$.

\begin{figure}[th]
\centerline
{\includegraphics[width=3in]{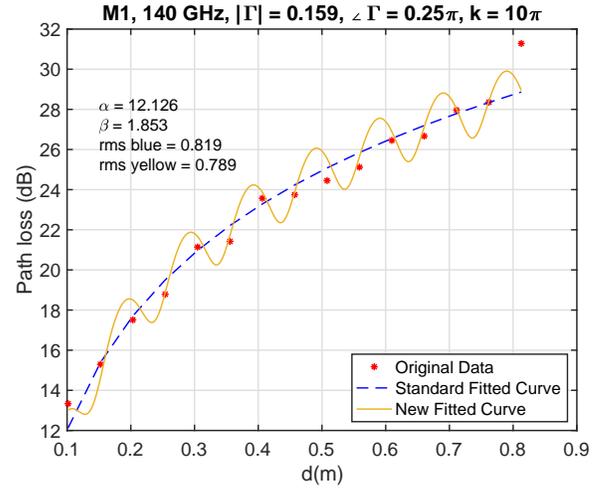}}
\caption{Result of manual fit for $\Gamma, k$ for 140 GHz.}
\label{140-manual}
\end{figure}

In equation \ref{vnet-amplitude} consider the term $\Re(\Gamma e^{i2k(d-d_0)})$. This is the real part of the complex term and can be represented as a cosine function of form $\cos(2k\pi d)$. By determining the distance between two consecutive maxima or minima in Figure \ref{140-standing}, we obtain the period of the cosine, from which we can then directly read off the value of $k$. In Figure \ref{140-standing}, this distance is 0.1m which gives an estimate of $k=10\pi$ for 140 GHz. Using these values for $\Gamma$ and $k$ we plot the overall fit in Figure \ref{140-manual}.

\begin{figure}[th]
\centerline
{\includegraphics[width=3in]{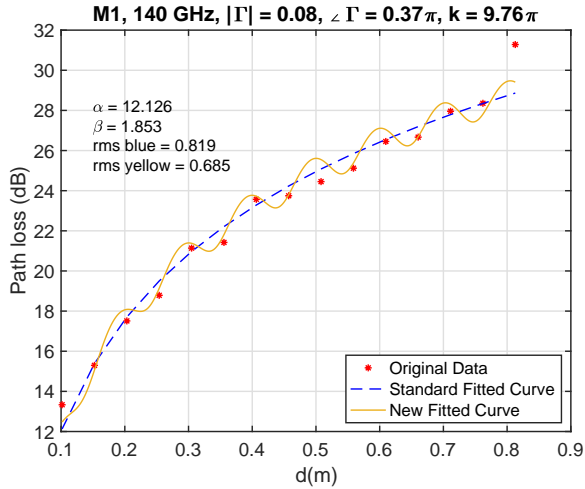}}
\caption{Result of rms fit for $\Gamma, k$ for 140 GHz.}
\label{140-rms}
\end{figure}

While this approach makes sense, there is one caveat -- since we only have a few measurements, we might have missed the correct values for $V_{max}$ and $V_{min}$ as well as the correct $k$. Therefore, we finally conduct a fine tuning step where we begin with the manually obtained values and do a rms fit varying $|\Gamma|, k$ and the phase of $\Gamma$. This fit gives us Figure \ref{140-rms} where we see an improved agreement with the measured data. The figure also shows the rms values for equation \ref{sffl} by itself and for our model which combines equation \ref{sffl} with \ref{vnet-amplitude}.

\begin{figure}[th]
\centerline
{\includegraphics[width=3in]{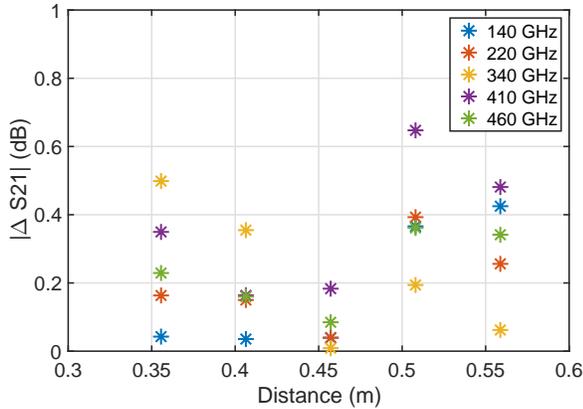}}
\caption{Difference in received signal between with and without wet towel}
\label{towel}
\end{figure}

\subsection{Model Fit for Different Frequencies} \label{linear-data}

We utilize the methodology outlined above to fit the measured data for the other four frequency bands. \textbf{To make sure that there is no significant reflection from the metallic surface of the measuring table, we compare the received signal in two cases: with and without a presence of a wet towel on top of the table which would absorb all reflections.} Figure \ref{towel} shows that there is less than 0.5 dB difference in the signal between the two cases. Therefore, we conclude that the effect of reflections off the table is minimal.

In every frequency, we observe that the {\em rms} error is smaller for the model that includes the standing wave equation. The values of the antenna reflection coefficient magnitude for 140, 220, 340, 410, 460 GHz are 0.08, 0.06, 0.05, 0.09, 0.0988 respectively. We can see that with an exception of 140 GHz data, as frequency gets higher, more energy is reflected back by the antenna. This implies a need for careful antenna selection when designing terahertz systems. It is also interesting to see that the value of $\beta$ in equation \ref{sffl} is 1.85 - 2.04 for the different frequency bands, giving us a simple propagation model (i.e., no reflected components with the exception of the standing wave).

\begin{figure}[th]
\centerline
{\includegraphics[width=3in]{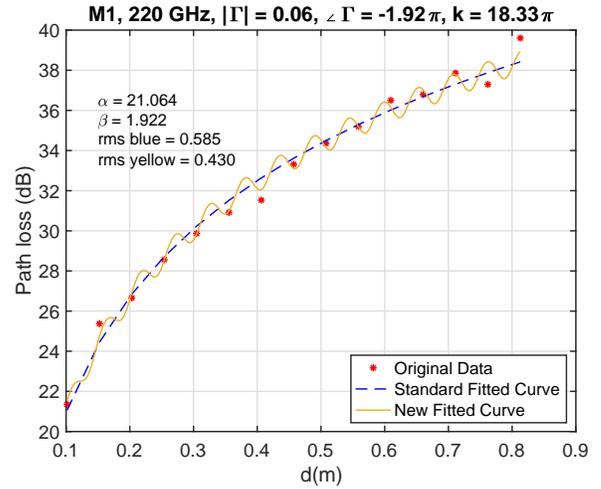}}
\caption{Fit for 220 GHz.}
\label{220-rms}
\end{figure}

\begin{figure}[th]
\centerline
{\includegraphics[width=3in]{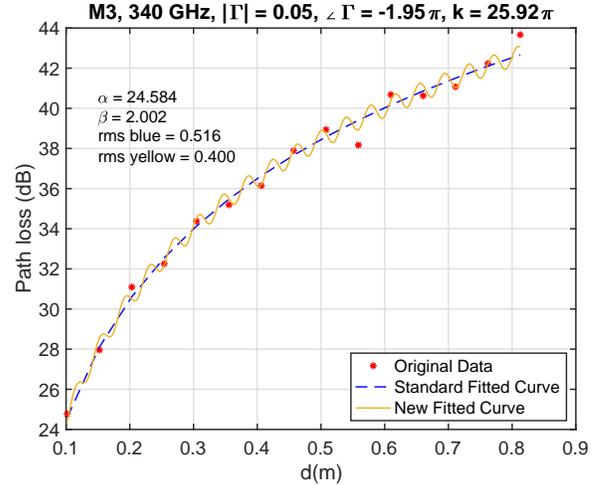}}
\caption{Fit for 340 GHz.}
\label{340-rms}
\end{figure}

\begin{figure}[th]
\centerline
{\includegraphics[width=3in]{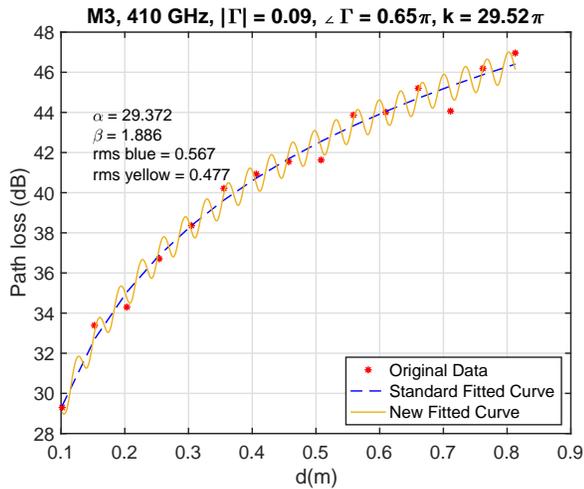}}
\caption{Fit for 410 GHz.}
\label{410-rms}
\end{figure}

\begin{figure}[th]
\centerline
{\includegraphics[width=3in]{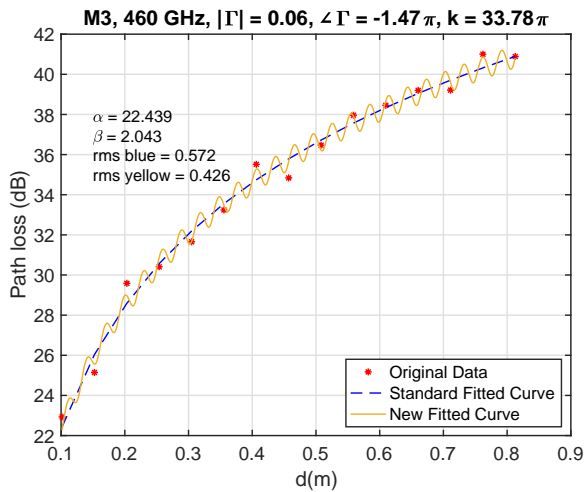}}
\caption{Fit for 460 GHz.}
\label{460-rms}
\end{figure}

\section{Conclusion} \label{conclude}

In this paper, we examine the propagation behavior of five terahertz frequency bands, 140, 220, 340, 410, 460 GHz, when using horn antennas at the transmitter and receiver. We modify existing channel models to include the effects of standing waves between the transmit and receive antennas. We show that at specific distances, these standing waves more than make up for the distance based attenuation. In other words, for certain distances and frequencies, as the distance increases, the attenuation falls. Our channel model provides an accurate fit for this behavior. 
For our future work, we will perform more measurements at finer distance granularity and examine other antenna types as well. We will also perform measurements in more complex propagation environments where there may be multiple reflectors in the path.

\bibliographystyle{unsrt}
\bibliography{reference,5g,thz,mynsfpubs}

\end{document}